# Diode-pumped mode-locked femtosecond 2-μm Tm:CaYAlO$_4$ laser

Lingchen Kong,[1] Guoqiang Xie,[1,4] Zhipeng Qin,[1] Xiaodong Xu,[2,5] Jun Xu,[3] Peng Yuan,[1] and Liejia Qian[1]

[1]Key Laboratory for Laser Plasmas (Ministry of Education), IFSA Collaborative Innovation Center, Department of Physics and Astronomy, Shanghai Jiao Tong University, Shanghai, 200240, China
[2]Jiangsu Key Laboratory of Advanced Laser Materials and Devices, School of Physics and Electronic Engineering, Jiangsu Normal University, Xuzhou 221116, China
[3]School of Physics Science and Engineering, Institute for Advanced Study, Tongji University, Shanghai 200092, China
[4]email: xiegq@sjtu.edu.cn
[5]email: xdxu79@mail.sic.ac.cn

**Abstract:** Continuous-wave mode-locked femtosecond 2-μm solid-state laser with a c-cut Tm:CaYAlO$_4$ (6 at.%) as gain medium was experimentally demonstrated. The mode locked laser generated stable pulses with average output power as high as 531 mW, pulse duration of 496 fs, and repetition rate of 97 MHz at 1975 nm. The research results show that Tm:CaYAlO$_4$ is an excellent gain medium for femtosecond pulse generation at 2 μm wavelength.

## 1. Introduction

Tm-ion or Ho-ion lasers at 2 μm wavelength are of great attraction for its applications in remote sensing [1], wind lidar [2, 3], laser processing [4], etc.. Since water exhibits a set of strong absorption lines near 2 μm, Tm-ion or Ho-ion lasers are very suitable for minimally invasive surgery [5]. Particularly, 2-μm femtosecond mode-locked lasers can provide ultrafast time resolution, wide comb spectrum, and high peak power, thus it can be used for ultrafast spectroscopy, gas sensing, mid-infrared supercontinuum generation and synchronous pumping of optical parametric oscillators for generation of mid and far infrared laser [6]. There is great motivation to search excellent laser materials for generation of 2-μm femtosecond laser [7-8]. So far, only several groups have reported on 2- μm femtosecond mode locking with Tm-doped crystals and ceramics [9-15].

Tm:CaYAlO4 (Tm:CaYAlO) is a disordered crystal with Tm dopants substituted for Y cation ions. The dopants, together with Ca$^{2+}$ and Y$^{3+}$ are distributed almost statistically between two equivalent large sites of A and B in nine coordinated sites of tetragonal ABCO$_4$ structure, leading to structural disorder and inhomogeneous broadening of optical transitions [16-19]. Thanks to Tm$^{3+}$ ions surrounded by CaYAlO tetragonal-symmetry-structured lattices, non-radiative relaxations of Tm:CaYAlO were quite weak [20, 21]. Tm:CaYAlO has attracted great interest as an excellent laser crystal because it possesses large absorption cross section, large emission cross section, and considerable cross-relaxation [22]. Meanwhile, CaYAlO host owns high mechanical strength, chemical stability and high thermal conductivity [22, 23]. In 1997, Moncorgke et al. reported on a Ti:sapphire laser-pumped continuous-wave Tm:CaYAlO laser with average output power of 50 mW [21]. Recently, diode-pumped continuous-wave (CW) Tm:CaYAlO laser with output power of 4.3 W and slope efficiency of 46.7% has been demonstrated in our group [24]. Wavelength tuning experiment shows that the Tm:CaYAlO laser can be continuously tuned from 1861 to 2046 nm [24]. These results indicate Tm:CaYAlO is a potential gain medium for generation of mode-locked femtosecond pulses with high output power. However, only picosecond mode-locked Tm:CaYAlO laser has been reported so far, which generated 35.3 ps pulses [25].

In this paper, we experimentally demonstrated a diode-pumped mode-locked femtosecond Tm:CaYAlO disordered crystal laser for the first time. By employing a semiconductor saturable absorber mirror (SESAM) with 3% modulation depth as mode locker, the Tm:CaYAlO oscillator emitted stable mode-locked pulses with pulse duration of 496 fs, repetition rate of 97 MHz and maximum average output power of 531 mW.

## 2. Experimental Setup

Figure.1 shows the schematic of the mode-locked femtosecond Tm:CaYAlO laser. The typical five-mirror X-folded cavity was adopted. A single-emitter AlGaAs laser diode at 790 nm was used as the pumping source. The divergent pumping beam was collimated and then focused into Tm:CaYAlO crystal by two coupling convex lenses L1 and L2. Both L1 and L2 had the same focus length of 100 mm. The focused pumping spot in the crystal was about 25 μm×86 μm in radius. The Tm:CaYAlO crystal employed in the experiment had a Tm concentration of 6 at.%, length of 9 mm and cross section of 4 mm × 4 mm. The sample was cut with Brewster angle to eliminate the Fresnel loss for p polarization light. The crystal was wrapped with indium foil and tightly mounted in a copper block cooled by circulating water. During the experiment, the circulating water temperature was set to 13 degrees. The three plano-concave mirrors M1, M2, and M3 had the same radius of curvature of -100 mm, and were all highly reflectively coated for laser wavelength (reflectivity >99.7% from 1850 nm to 2100 nm) and anti-reflectively coated for pumping wavelength (transmission>95% at 790 nm). A pair of CaF$_2$ prism with tip to tip distance of 35 cm was inserted into the optical path to compensate the dispersion of the resonator. The commercial SESAM (BATOP, SAM-1960-5-10ps-x) operating at 2 μm wavelength had a modulation depth of 3% and saturation energy fluence of 30 μJ/cm$^2$. The wedged plano-plano output coupler (OC) had a transmission of 2%. The laser mode sizes in the crystal and on the SESAM were calculated to be about 35 μm × 33 μm and 77 μm × 80 μm in radius, respectively.

## 3. Experimental Results and Discussion

*3.1 CW operation:*

First, a plano-plano high-reflectivity mirror displaced the SESAM in the cavity for testing CW laser characteristics of Tm:CaYAlO. The laser output power was measured with a thermal powermeter (FieldMate, Coherent Co.). The CW laser output result is shown in Fig. 2. The laser began to oscillate at the pumping threshold of 0.531 W. The laser output power linearly increased as the pump power, with a slope efficiency of 33.9%. A maximum CW output power of 1.59 W was obtained at the absorbed pumping power of 5.34 W. Higher output power was only limited by the available pump power.

*3.2 CW mode locking operation:*

With SESAM as mode locker, stable continuous wave (CW) mode locking was established under an absorbed pumping power of 2.96 W. The pulse trains were monitored by a 500-MHz-bandwidth oscilloscope (Tektronix, DPO3054) and a high-speed detector (EOT, ET-5000). The mode locked pulse train is shown in Fig. 3. The mode locking had a good stability and could be sustained for hours. The radio-frequency (RF) spectrum of mode locking was shown in Fig. 4, measured by a 26.5 GHz microwave spectrum analyzer (N9938A, FieldFox). The RF spectrum shows a signal-to-noise ratio of 51 dB, indicating a clean mode locking. The mode locked pulses had a repetition rate of 97 MHz according to RF spectrum, corresponding to a cavity length of 153 cm. In CW mode locking operation, the average output power increased with the pumping power, and a maximum average output power of 531 mW was obtained under an absorbed pumping power of 5.04 W (Fig. 2). The output beam had a round $TEM_{00}$ mode.

The mode-locked pulse spectrum was measured with a mid-infrared (mid-IR) optical spectrum analyzer (Ocean Optics, SIR5000) with a resolution of 0.22 nm, as shown in Fig. 5. The spectrum had a FWHM bandwidth of 9.8 nm centered at 1975 nm. With a commercial autocorrelator (APE, PulseCheck50), the autocorrelation trace of the mode locked pulses was recorded (Fig. 6). The mode locked pulses had a pulse duration of 496 fs, assuming a $sech^2$ pulse shape. The time-bandwidth product was calculated to be 0.37, which was close to the Fourier transform limit value.

## 4. Conclusion

By adopting SESAM with 3% modulation depth, stable femtosecond mode-locking of Tm:CaYAlO$_4$ laser has been realized. The mode-locked laser generated pulses with pulse duration of 496 fs, average output power of 531 mW, repetition rate of 97 MHz, and pulse energy of 5.5 nJ. The high-power femtosecond laser at 2 μm wavelength is expected to have important applications in ultrafast molecules spectroscopy, generation of mid-infrared supercontinuum, and synchronously pumping of mid-infrared OPO, etc.

**Acknowledgements**

The work is partially supported by the Shanghai Excellent Academic Leader Project (Grant No. 15XD1502100), National Natural Science Foundation of China (Grant No. 11421064) and the National Basic Research Program of China (Grant No. 2013CBA01505).

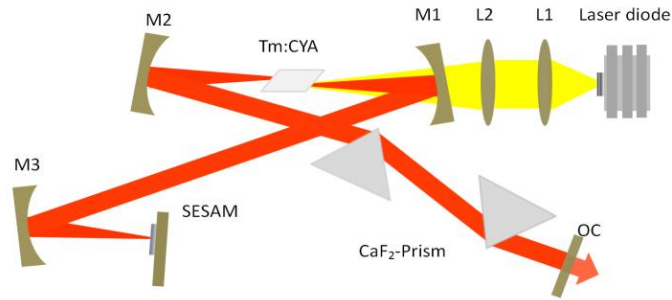

Fig.1 Schematic of the mode-locked femtosecond Tm:CaYAlO$_4$ laser. L1, L2: doublet lens with the same focal length of 100 mm; M1, M2 and M3: plano-concave mirrors with radius of curvature (ROC) of -100mm; OC: output coupler. A pair of CaF2 prisms was inserted for dispersion compensation.

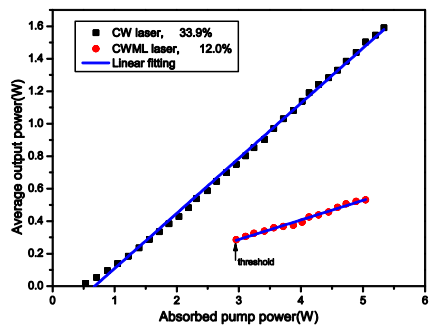

Fig. 2 Average output power versus absorbed pumping power in CW and CW mode locking (CWML) regimes.

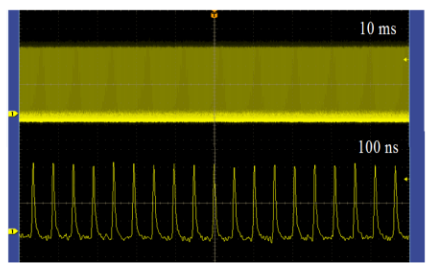

Fig. 3 Mode-locked pulse trains in 10 ms and 100 ns time scales, respectively.

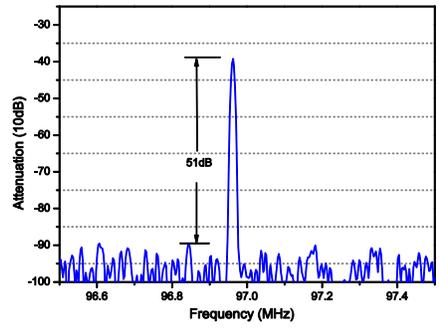

Fig. 4 Fundamental radio frequency spectrum of mode locked pulses.

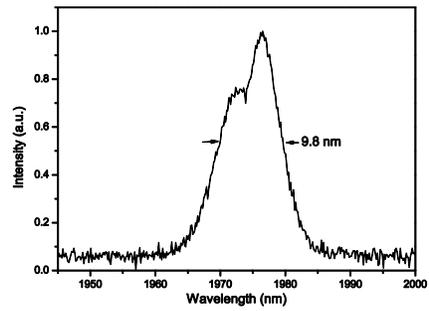

Fig. 5 The optical spectrum of femtosecond mode-locked Tm:CaYAlO$_4$ laser

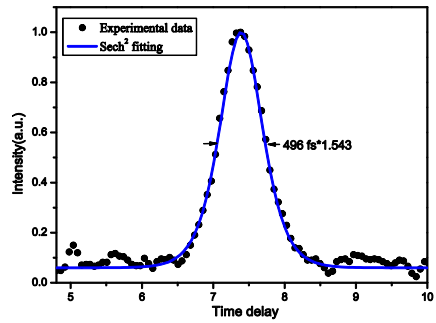

Fig. 6 the autocorrelation trace of the mode locked pulses.